\documentclass[aps, floatfix,twocolumn,preprintnumbers, superscriptaddress]{revtex4-1}
\usepackage{hyperref}
\usepackage{verbatim}
\usepackage{amsmath}
\usepackage{latexsym}
\usepackage{natbib}
\usepackage{amsmath}
\usepackage{amssymb}
\usepackage{amsthm}
\usepackage{graphicx}
\usepackage{bm}
\usepackage{color}

\newcommand{\ra}{\rightarrow}

\newcommand{\m}[1]{\mathrm{#1}}

\usepackage{amsfonts}

\newcommand{\px}{\sigma^x}
\newcommand{\py}{\sigma^y}
\newcommand{\pz}{\sigma^z}

\usepackage{pict2e} 
\usepackage{rotating} 

\begin{document}

\title{An efficient Markov chain Monte Carlo algorithm for the surface code}
\author{Adrian Hutter}
\author{James R.~Wootton}
\author{Daniel Loss}
\affiliation{Department of Physics, University of Basel, Klingelbergstrasse 82, CH-4056 Basel, Switzerland}

\date{\today}

\begin{abstract}
Minimum-weight perfect matching (MWPM) has been been the primary classical algorithm for error correction in the surface code, since it is of low runtime complexity and achieves relatively low logical error rates [Phys.\ Rev.\ Lett.\ \textbf{108}, 180501 (2012)]. 
A Markov chain Monte Carlo (MCMC) algorithm [Phys.\ Rev.\ Lett.\ \textbf{109}, 160503 (2012)] is able to achieve lower logical error rates and higher thresholds than MWPM, 
but requires a classical runtime complexity which is super-polynomial in $L$, the linear size of the code. 
In this work we present an MCMC algorithm that achieves significantly lower logical error rates than MWPM at the cost of a polynomially increased classical runtime complexity.
For error rates $p$ close to the threshold, our algorithm needs a runtime complexity which is increased by $O(L^2)$ relative to MWPM in order to achieve a lower logical error rate. 
If $p$ is below an $L$-dependent critical value, no increase in the runtime complexity is necessary any longer.
For $p\ra0$, the logical error rate achieved by our algorithm is exponentially smaller (in $L$) than that of MWPM, without requiring an increased runtime complexity.
Our algorithm allows for trade-offs between runtime and achieved logical error rates as well as for parallelization, and can be also used to correct in the case of imperfect stabilizer measurements.

\end{abstract}
\maketitle

\section{Introduction}

An important primitive for the \emph{processing} of quantum information is the ability to \emph{store} it despite constant corruptive influence of the external environment on the applied hardware and imperfections of the latter.
While one approach seeks to achieve this by constructing a \emph{self-correcting quantum memory} (see Ref.~\cite{Woot12b} for a recent review), an alternative possibility is to dynamically protect the stored quantum information by constantly pumping entropy out of the system.
Topological quantum error correction codes \cite{Kitaev03,Freed02} store one logical qubit in a large number of physical qubits, in a way which guarantees that a sufficiently low density of errors on the physical qubits can be detected and undone, without affecting the stored logical qubit.
Most promising is the surface code \cite{Brav98,Denn02,Raus07,Fowl08}, which requires only local four-qubit parity operators to be measured. 
While proposals for direct measurement of such operators exist \cite{Pedro11,DiVi12,Nigg12}, most of the literature focuses on time-dependent interactions between the four qubits and an auxiliary qubit, allowing to perform sequential CNOT gates and to finally read the measurement result off the auxiliary qubit.
See Ref.~\cite{Fowl12a} for a recent review. 

In order to decode the syndrome information, i.e., use the outcomes of all four-qubit measurements to find out how to optimally perform error correction, a classical computation is necessary.
This classical computation is not trivial and brute force approaches are infeasible.
Decoding algorithms based on renormalization techniques \cite{Ducl09} or minimum-weight perfect matching (MWPM) \cite{Fowl12b} have a runtime complexity $O(L^2)$ and can be parallelized to $O(L^0)$ (neglecting logarithms), where $L$ is the linear size of the code.
As these algorithms are approximative, the logical error rates achievable with them fall short of those theoretically achievable by brute force decoding. 
A Markov chain Monte Carlo (MCMC) algorithm \cite{Woot12a} can cope with higher physical error rates than the two mentioned algorithms, but has super-polynomial (yet sub-exponential) runtime complexity.
In this work, we present an efficient MCMC decoding algorithm that allows to achieve logical error rates lower than those achievable by means of MWPM \cite{comparison_comment}.
Equivalently, a smaller code size is required to achieve a certain target logical error rate. 
Our algorithm allows for trade-offs between runtime and achieved logical error rate. 
If we define the runtime or our algorithm to be the minimal computation time such that the achieved logical error rate is lower than the one achievable by means of MWPM, we find it to be at most $O(L^4)$.
The runtime complexity can be parallelized to $O(L^2)$, thus adding complexity $O(L^{2})$ to MWPM and renormalization technique decoding algorithms.
For low enough error rates, it is found that even an $O(1)$ increase in the runtime relative to MWPM allows us to achieve a lower logical error rate.

In summary, in comparison to alternative algorithms \cite{Ducl09,Fowl12b} our algorithm allows for lower quantum information error rates and smaller code sizes at the cost of a (polynomially) higher classical runtime complexity. 
Given the current state of the art of quantum and classical information processing, shifting requirements from quantum to classical seems desirable.  
Our algorithm is generalizable to the (realistic) case of imperfect stabilizer measurements, though we restrict numerical simulations in this work to the case of perfect measurements for simplicity.

\section{Error correction in surface codes}

\begin{figure}
	\includegraphics[width=0.5\textwidth]{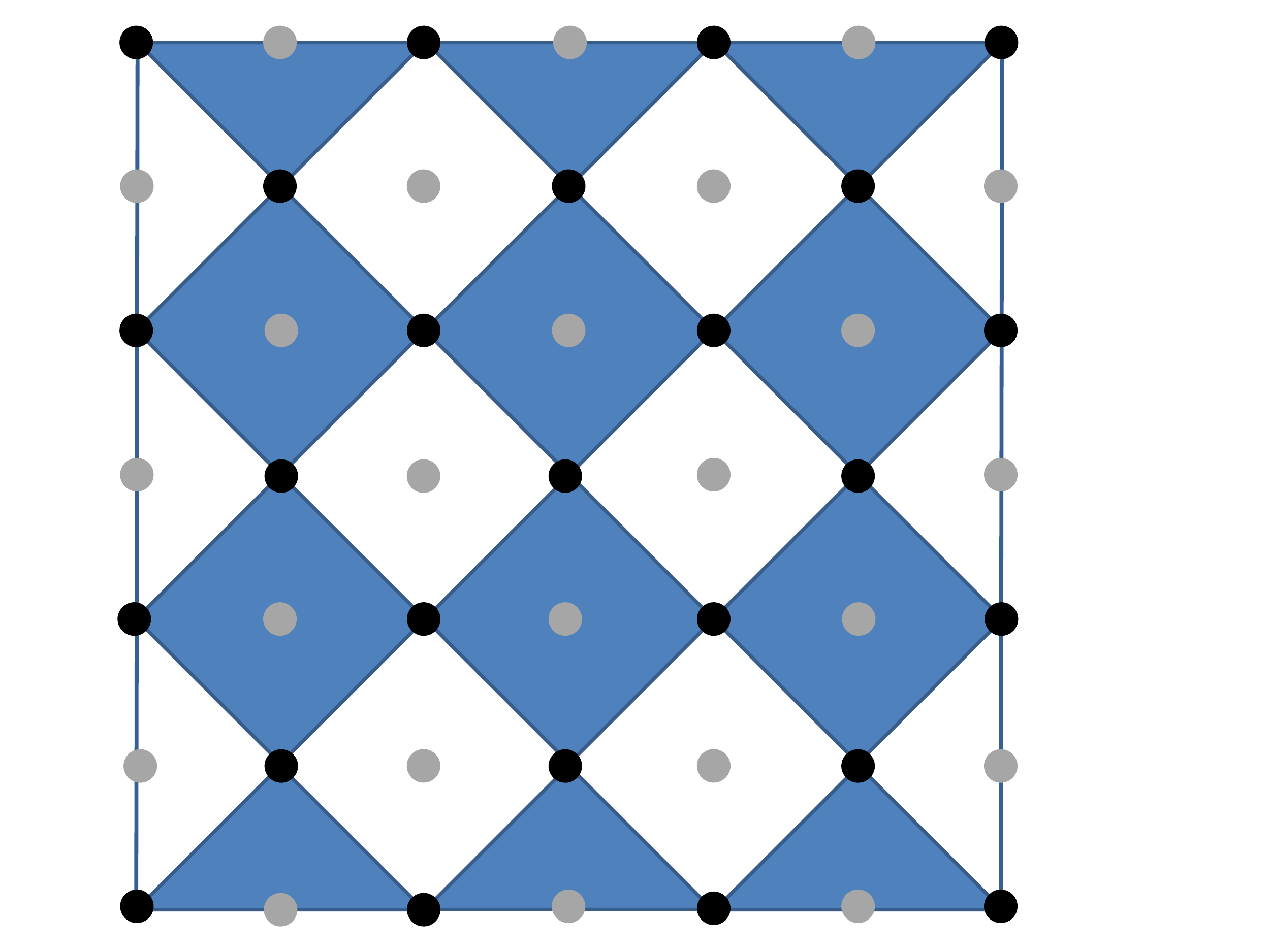}
	\caption{An $L=4$ surface code. Black dots are data qubits, grey dots are syndrome qubits that allow to read off the results of the stabilizer measurements when sequential CNOT gates have been performed between them and the adjacent data qubits.
	Stabilizer operators are either tensor products of $\px$ operators (acting on the data qubits around a white square/triangle) or tensor products of $\pz$ operators (acting on the data qubits around a blue square/triangle).}
	\label{fig:code}
\end{figure}

Stabilizer operators are, in the context of the surface code, tensor products of $\px$ or $\pz$ operators (see Fig.~\ref{fig:code}) which are required to yield a $+1$ eigenvalue when applied to the quantum state stored in the code.
Eigenvalues $-1$ are treated as errors and interpreted as the presence of an \emph{anyon}. A surface code of size $L$ has $n_{\text{stab}}=2L(L-1)$ ($3$- and $4$-qubit) stabilizers. 
Since all stabilizers commute, they can be measured simultaneously and hence the presence of anyons can be detected.
Any Pauli operator $\px$, $\py$, or $\pz$ applied to a data qubit creates at least one anyon as it anti-commutes with at least one stabilizer.
We call violated $\px$-stabilizers s-anyons and violated $\pz$-stabilizers p-anyons. 

Given some anyon configuration $A$, the goal is to apply a series of single-qubit $\px$ and $\pz$ operators, such that all anyons are removed and a trivial operation has been performed on the code subspace. 
Two such hypotheses about what errors the physical qubits have suffered are equivalent if they can be deformed into each other through the application of stabilizers. 
Equivalent error chains will lead to the same operation performed on the code subspace (consisting of the states which are $+1$ eigenstates of each stabilizer). 
For the surface code, there are four such equivalence classes. The goal is therefore to find the most probable equivalence class of error chains and not to find the most likely error chain.
The most likely error chain need not be an element of the most likely equivalence class, though trying to correct by undoing the most likely error path is a reasonable approximation and is the idea behind minimum-weight perfect matching correction algorithms.
More precisely, MWPM matches both kinds of anyons independently of each other and thus ignores potential correlations between $\px$- and $\pz$-errors.

Decoherence models in which each qubit independently is subject to the channel
\begin{align}\label{eq:pauliChannel}
 \rho \mapsto p_I\rho + p_x\px\rho\px + p_y\py\rho\py + p_z\pz\rho\pz  
\end{align}
(with $p_I+p_x+p_y+p_z=1$) allow for efficient simulation on a classical computer.
While physical decoherence models may not exactly have the form of Eq.~(\ref{eq:pauliChannel}), they may be approximated by such a channel through a \emph{Pauli twirl approximation} \cite{Dur05,Emer07}.
The two most frequently studied noise models of the form of Eq.~(\ref{eq:pauliChannel}) are \emph{independent bit- and phase-flip errors} ($p_x = p_b(1-p_p)$, $p_z = p_p(1-p_b)$ and $p_y=p_b p_p$ for independent bit- and phase-flip probabilities $p_b$ and $p_p$) and \emph{depolarizing noise} ($p_x=p_z=p_y=\frac{p}{3}$).
The theoretical maximal error rates up to which error correction is possible by exact error correction are known to be $p_b,p_p<10.9\%$ for independent bit- and phase-flip errors \cite{Denn02} and $p<18.9\%$ for depolarizing noise \cite{Bombin12}.
Any approximate error correction algorithm will yield threshold error rates below these theoretical maxima.

Minimum weight matching considers bit flip errors (which create $p$-anyons) and phase flip errors (which create $s$-anyons) independently. As such it is only well designed for noise models with no correlations between $\px$- and $\pz$-errors. Errors models that do have these correlations, such as depolarizing noise, can only be treated approximately. Typically this means that the correction will be done as if the bit and phase flip errors occurred with independent probabilities $p_b = p_x + p_y$ and $p_p = p_x + p_y$, calculated according to the true (correlated) noise model.

This suboptimal treatment of correlated noise leads to suboptimal behaviour. Thresholds are significantly lower than the theoretical maxmima and the effectiveness below threshold is also significantly affected. For example, let us consider the behaviour for depolarizing noise with very low $p$. In this case the probability of a logical error is dominated by the probability of the most likely fatal error pattern (one which causes the decoder to guess the wrong equivalence class). This means the fatal error pattern with the minimum number of single qubit errors.

Perfect matching treats this noise model as one in which bit and phase flip errors occur independently with probabilities $p_b = p_p = 2p/3$. A fatal error pattern that causes a logical bit flip error requires, for odd $L$, at least $\frac{L+1}{2}$ single qubit bit flips to occur in a line, such that they create a pair of $p$-anyons separated by just over half the size of the code (or a single $p$-anyon which is closer to the boundary to which it is not connected by the error chain) . This is because the matching  will incorrectly think that they were created by the $\frac{L-1}{2}$ bit flips required to create them within the opposite equivalence class, since this matching has a smaller weight. An equivalent argument holds for logical phase flips. The probability of logical errors in the low $p$ limit is then $\sim \max \left( p_b^{\frac{L+1}{2}}, p_p^{\frac{L+1}{2}} \right) = (2p/3)^{\frac{L+1}{2}}$.

An optimal decoder will behave slightly differently, since it can see the difference between bit-flips caused by $\px$ errors and those caused by $\py$. A line of $\frac{L+1}{2}$ bit flips is then only sufficient to cause a logical bit-flip error if it contains at most one error caused by $\py$, since the presence of  $\py$'s in the line will create $s$-anyons along its length. This leaves a trail of breadcrumbs to show the decoder that the larger weight matching is in fact the correct one. An equivalent argument holds for logical phase flips. 
It is therefore a chain of $\frac{L+1}{2}$ errors with at most one $\py$ error and all remaining ones either $\px$ or $\pz$ errors that are required for a logical error.
The probability of these in the low $p$ limit is then $\sim \max \left( p_x^{\frac{L-1}{2}+O(\log L)}, p_z^{\frac{L-1}{2}+O(\log L)} \right) = (p/3)^{\frac{L-1}{2}+O(\log L)}$.
Clearly, a decoder which is able to take correlations between bit- and phase-flips into account will thus for $p\ra0$ lead to a logical error rate which is smaller by a factor $O(2^{\frac{L-1}{2}})$ than the one for standard MWPM.

In order to address this issue, we introduce two new algorithms to perform decoding.
The first, an enhanced version of MWPM, is discussed in Sec.~\ref{sec:enhanced}. It performs optimally in the limit of $p\ra0$ and is also able to outperform standard MWPM for non-vanishing values of $p$.
It is based on performing MWPM for four sligtly modified anyon configurations and thus does not require an enhanced runtime complexity relative to MWPM.
The second algorithm, discussed in Sec.~\ref{sec:MCMC}, uses enhanced MWPM as a starting point and then performs Markov chain Monte Carlo sampling in order to further reduce the logical error rate.

This requires a runtime complexity that is at most an $O(L^2)$ factor greater than that of MWPM. However, even this modest increase is only required for $p$ close to threshold. For lower values of $p$, no increase in the runtime complexity is necessary \cite{log_comment}.

\section{Enhanced MWPM} \label{sec:enhanced}

Our first method consists of only a small change to the standard MWPM decoding, but it nevertheless has a large effect. To explain this fully, we must first explain standard MWPM in more detail.

For a graph with weighted edges between an even number of vertices, Edmond's MWPM algorithm \cite{Edmo65} finds the pairing of minimal weight efficiently.
We employ the library \texttt{Blossom V} \cite{Kolmogorov09} to perform MWPM. For the graphs which are relevant for the surface code, MWPM can be performed in runtime complexity $O(L^2)$ and can be parallelized to $O(L^0)$ \cite{Fowl12b}.
In order to obtain such graphs, we assign to the edge between two anyons (corresponding to the vertices) the minimal number of single-qubit errors necessary to link them (their Manhattan distance).
For each anyon, we place a virtual partner on the closer boundary of the type which is able to absorb it (top an bottom for s-anyons and left and right for p-anyons). 
We then add an edge between each anyon and its virtual partner (with weight again given by the Manhattan distance) and zero-weight edges between all virtual anyons on the same boundary.
Including these virtual anyons ensures that the number of vertices in the graph is even and that each anyon can be matched to the closest absorbing boundary. The zero-weight edges ensure that unnecessary virtual anyons can be removed at no cost.
The four equivalence classes of errors in the surface code may be identified by determining the parity of the number of errors that lie on a given line that links the top and bottom or left and right boundary.

\begin{figure}
  \setlength{\unitlength}{0.5\textwidth}
  \begin{picture}(0.8,0.5)
	\put(-0.15,0.075){\includegraphics[width=0.25\textwidth]{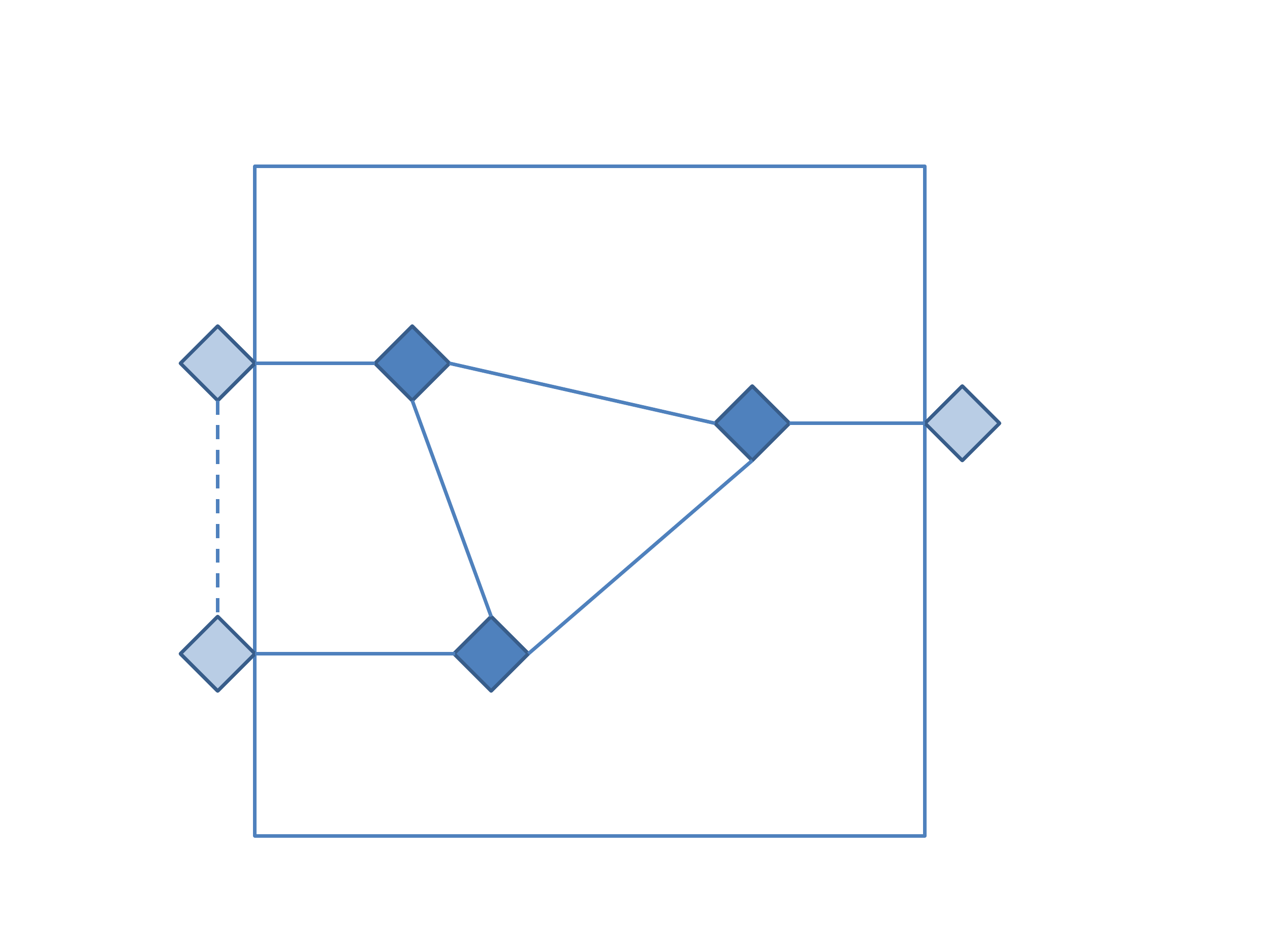}}
	\put(0.35,0.04){\includegraphics[width=0.30\textwidth]{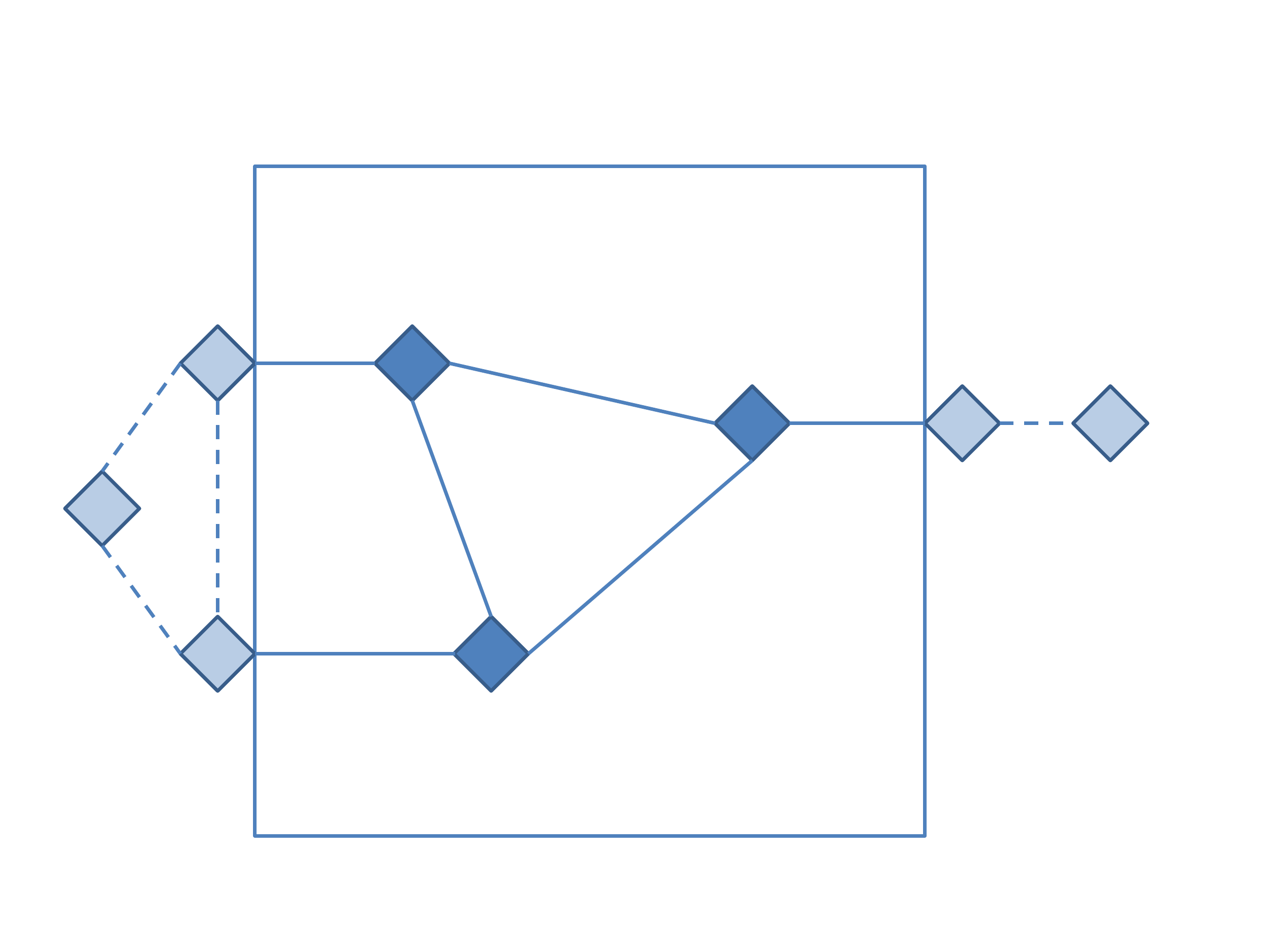}}
	\put(-0.09,0.5){\Large a)}
	\put(0.43,0.5){\Large b)}
  \end{picture}
  \caption{Three p-anyons (solid squares) have been detected in a surface code. In order to find hypotheses of minimum weight (maximal probability) about what errors have occurred, 
we first add virtual anyons (light squares) on the closest absorbing boundary of each real anyon and connect virtual anyons residing on the same boundary; see the a) part of the figure. 
Dashed lines represent zero-weight edges, solid lines represent non-zero-weight edges. In part b) of the figure, we place an additional virtual anyon on the left and right boundary.
Note that each possible pairing in part~b) is an element of a different equivalence class then the pairings which are possible in part~a).}
  \label{fig:equivClasses}
\end{figure}

This method outputs only a single matching for a single equivalence class, that which yields the minimum weight matching overall. To make our enhancement we will force the decoder to output the minimum weight matching for each equivalence class, which requires us to enforce changes in the parities of error numbers along lines across the code. This can be achieved by adding one further virtual anyon on the top and bottom or the left and right boundary, respectively, and connecting it with zero-weight edges to all virtual anyons already present at this boundary \cite{no_anyons_there_comment}.
It is thus guaranteed that a any pairing of all real and virtual anyons of the same type is an element of a different equivalence class than the one obtained if these two additional virtual anyons had not been included.
See Fig.~\ref{fig:equivClasses} for an illustration.

This prodcedure allows us to find two non-equivalent minimum-weight error chains for both kinds of anyons. 
Combining these $2\times2$ error chains gives four hypotheses about the errors that have happened. Each of these matchings is the most likely within its equivalence class for the approximate error model where the correlations between $x$- and $z$-errors are ignored.  , we may now determine which is most likely according to the true correlated noise model, by simply determining the probability for the total error chain in each. For the case of depolarizing noise this means determining which has the minimal number of errors, where one $x$- and one $z$-error on the same qubit count only as one $y$-error (rather than two errors, as counted by standard matching). 

\begin{figure}
  \setlength{\unitlength}{0.5\textwidth}
  \begin{picture}(0.8,0.6)
	\put(-0.12,-0.02){\includegraphics[width=0.50\textwidth]{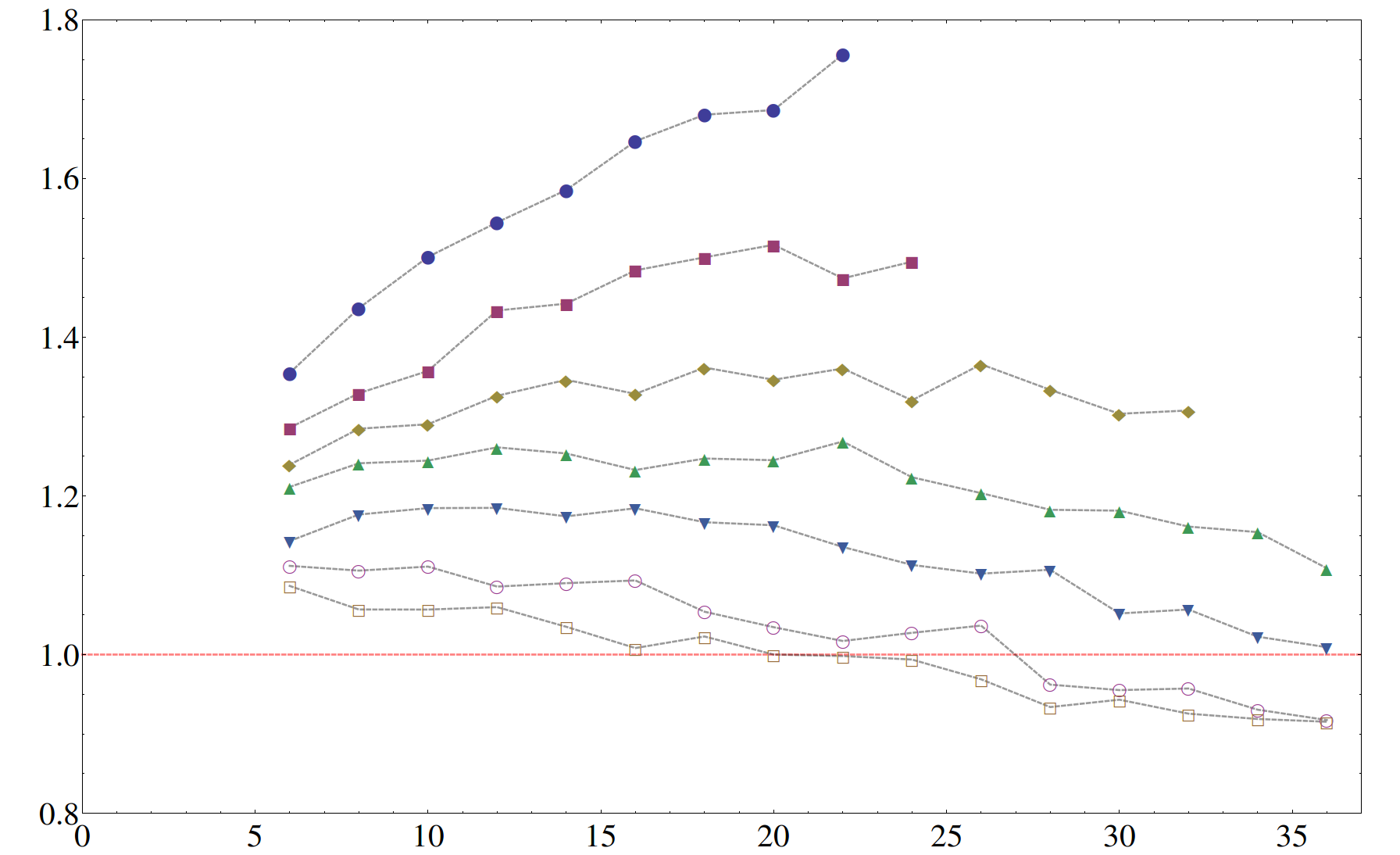}}

	\put(0.49,0.55){\tiny $p=6\%$}
	\put(0.54,0.40){\tiny $p=7\%$}
	\put(0.74,0.29){\tiny $p=8\%$}
	\put(0.74,0.22){\tiny $p=9\%$}
	\put(0.74,0.16){\tiny $p=10\%$}
	\put(0.74,0.10){\tiny $p=12\%$}
	\put(0.74,0.05){\tiny $p=14\%$}

  \end{picture}
  \caption{Ratio of the logical error rate of standard MWPM divided by the logical error rate of enhanced MWPM for various codes sizes $L$ (horizontal axis) and error rates $p$. Data points below the red line indicated that enhanced MWPM performs worse than standard MWPM. Each data point is sampled over as many error configurations as were necessary in order to obtain $10\,000$ logical errors. The grey dashed lines are guides to the eye.}
  \label{fig:MWPM}
\end{figure}
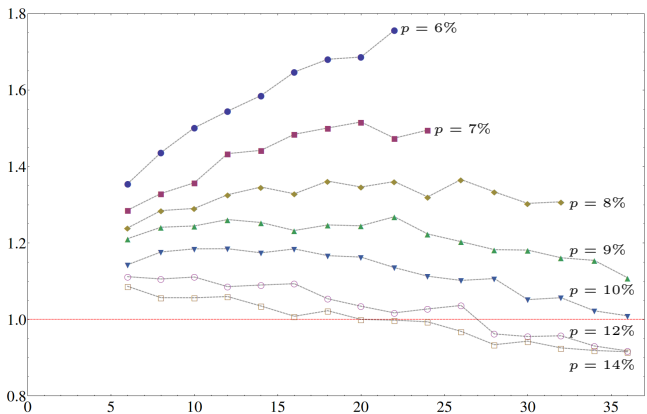

Fig.~\ref{fig:MWPM} compares the logical error rates of enhanced and standard MWPM for error rates $p$ which are of the same order of magnitude as the threshold value.
Very surprisingly, there are regimes (both $L$ and $p$ large enough), where ``enhanced'' MWPM performs worse than standard MWPM.
While we know that for low enough $p$ the ratio displayed in Fig.~\ref{fig:MWPM} increases exponentially with $L$, the ratio \emph{decreases} with $L$ for $p\gtrsim8\%$. 
We can offer no explanation for this behavior. Further comparison of enhanced and standard MWPM can be found in Fig.~\ref{fig:differentAlgos}.

\section{Markov chain Monte Carlo Algorithm}\label{sec:MCMC}
 
We now consider an algorithm based on an analytically exact rewriting of the probability of each equivalence class which allows evaluation with the Metropolis algorithm. 
Let us discuss depolarizing noise here and note that our discussion generalizes straightforwardly to arbitrary error models of the form of Eq.~(\ref{eq:pauliChannel}).
We have defined a depolarization rate $p$ to mean that each spin has suffered a $\px$, $\py$, or $\pz$ error with probability $p/3$ each and no error with probability $1-p$.
Consequently, the probability of an error chain involving $n$ single-qubit errors is up to a normalization constant given by $\left(\frac{p/3}{1-p}\right)^n\equiv e^{-\bar{\beta} n}$, where $\bar{\beta}$ is defined through 
\begin{align}\label{eq:betaBar}
 \bar{\beta}=-\log\left(\frac{p/3}{1-p}\right)\ .
\end{align}
Given an anyon configuration $A$, the relative probability of equivalence class $E$ can be written as
\begin{align}\label{eq:partitionFunction}
 Z_E(\bar{\beta}) = \sum_E e^{-\bar{\beta} n}\ ,
\end{align}
where the sum runs over all error chains that are compatible with the anyon configuration $A$ and elements of equivalence class $E$, and $n$ denotes the number of single-qubit errors in a particular error chain.
The goal is to find the equivalence class $E$ with maximal $Z_E(\bar{\beta})$. 

The Metropolis algorithm allows us to approximate expressions of the form
\begin{align}\label{eq:avgDefi}
 \left\langle f(n) \right\rangle_{\beta, E} := \frac{\sum_E f(n) e^{-\beta n}}{Z_E(\beta)}
\end{align}
(we use $\beta$ to denote a generic ``inverse temperature'' and $\bar{\beta}$ to denote the specific one defined through Eq.~(\ref{eq:betaBar})).
The sum is here over all error configurations in equivalence class $E$ that are compatible with the syndrome information $A$. 
In order to approximate an expression of the form in Eq.~(\ref{eq:avgDefi}) by use of the Metropolis algorithm, we pick one stabilizer at random and calculate the number $\Delta n$ by which the total number of errors $n$ in the code would change if that stabilizer were applied.
If $\Delta n\leq0$, we apply the stabilizer and if $\Delta n>0$ we apply it with probability $e^{-\beta \Delta n}$. Summing up $f(n)$ over all steps and dividing by the total number of steps then yields our approximation to Eq.~(\ref{eq:avgDefi}).

Deforming error patterns only through the application of stabilizers ensures that all error patterns in one such Markov chain belong to the same class, and that all of them are compatible with the same anyon configuration $A$. 
Since we will need the average $\left\langle f(n) \right\rangle_{\beta, E}$ for each equivalence class $E$, we need an initial error configuration from each equivalence class which is compatible with the measured anyon syndrome $A$. In fact, we will start the Metropolis Markov chains with the minimum weight error configuration from each equivalence class, provided by the method described above in Section \ref{sec:enhanced} and Fig. \ref{fig:equivClasses}. The reason for starting with the minimum weight error configuration rather than a random initial configuration from the same equivalence class is based on the intuition that ``heating up'' from the groundstate to inverse temperatures $\beta$ as needed for the equilibrium averages in Eq.~(\ref{eq:avgDefi}) takes less time than ``cooling down'' from a high energy configuration. 

Note that $\sum_E$ in Eq.~(\ref{eq:avgDefi}) has $2^{n_{\text{stab}}}$ summands for each equivalence class $E$, so knowing an averaged sum is as good as knowing the whole sum.
We have
\begin{align}
 Z_E(\bar{\beta}) = \left\langle e^{-\bar{\beta}n} \right\rangle_{\beta=0, E}\times 2^{n_{\text{stab}}}\ ,
\end{align}
corresponding to a simple Monte Carlo sampling of the sum.
However, the sum is dominated by an exponentially small fraction of summands with ``energy'' $n$ close to the minimal value, so Monte Carlo sampling is computationally similarly expensive as a brute force calculation of the sum.
Our goal is thus to rewrite $Z_E(\bar{\beta})$ in a way that involves only quantities which are evaluable efficiently with the Metropolis algorithm.
Applying the fundamental theorem of calculus we have
\begin{align}\label{eq:main}
 \log Z_E(\bar{\beta}) &= \int_0^{\bar{\beta}}\m{d}\beta\partial_\beta \log Z_E(\beta) + \log Z_E(\beta=0) \nonumber\\
&= -\int_0^{\bar{\beta}}\m{d}\beta\left\langle n \right\rangle_{\beta, E} + n_{\text{stab}}\log2\ .
\end{align}
If we know the functions $\left\langle n \right\rangle_{\beta, E}$, the most likely equivalence class is, according to Eq.~(\ref{eq:main}), the one in which the area under the curve is smallest.
In the correctable regime ($p<p_c$) the differences in ``free energy'' 
\begin{align}\label{eq:freeEn}
 F_E(\bar{\beta}) &= -\frac{1}{\bar{\beta}}\log Z_E(\bar{\beta}) \nonumber\\
 &= \frac{1}{\bar{\beta}}\int_0^{\bar{\beta}}\m{d}\beta\left\langle n \right\rangle_{\beta, E} + \m{const}
\end{align}
between the different equivalence classes grow proportionally in $L$ and correspondingly the probability of all equivalence classes but the most likely one decreases exponentially with $L$.

For a positive $\beta$, the average number of errors $\left\langle n \right\rangle_{\beta, E}$ can be efficiently calculated to arbitrary accuracy by means of the Metropolis algorithm.
The integral $\int_0^{\bar{\beta}}\m{d}\beta$ can be calculated to arbitrary accuracy by first calculating the values $\left\langle n \right\rangle_{\beta, E}$ for a sufficient number of inverse temperatures $\beta$ and then applying a quadrature formula like Simpson's Rule.

\subsection{The single-temperature algorithm}

\begin{figure}
  \setlength{\unitlength}{0.5\textwidth}
  \begin{picture}(0.72,0.5)
	\put(-0.12,-0.15){\includegraphics[width=0.50\textwidth]{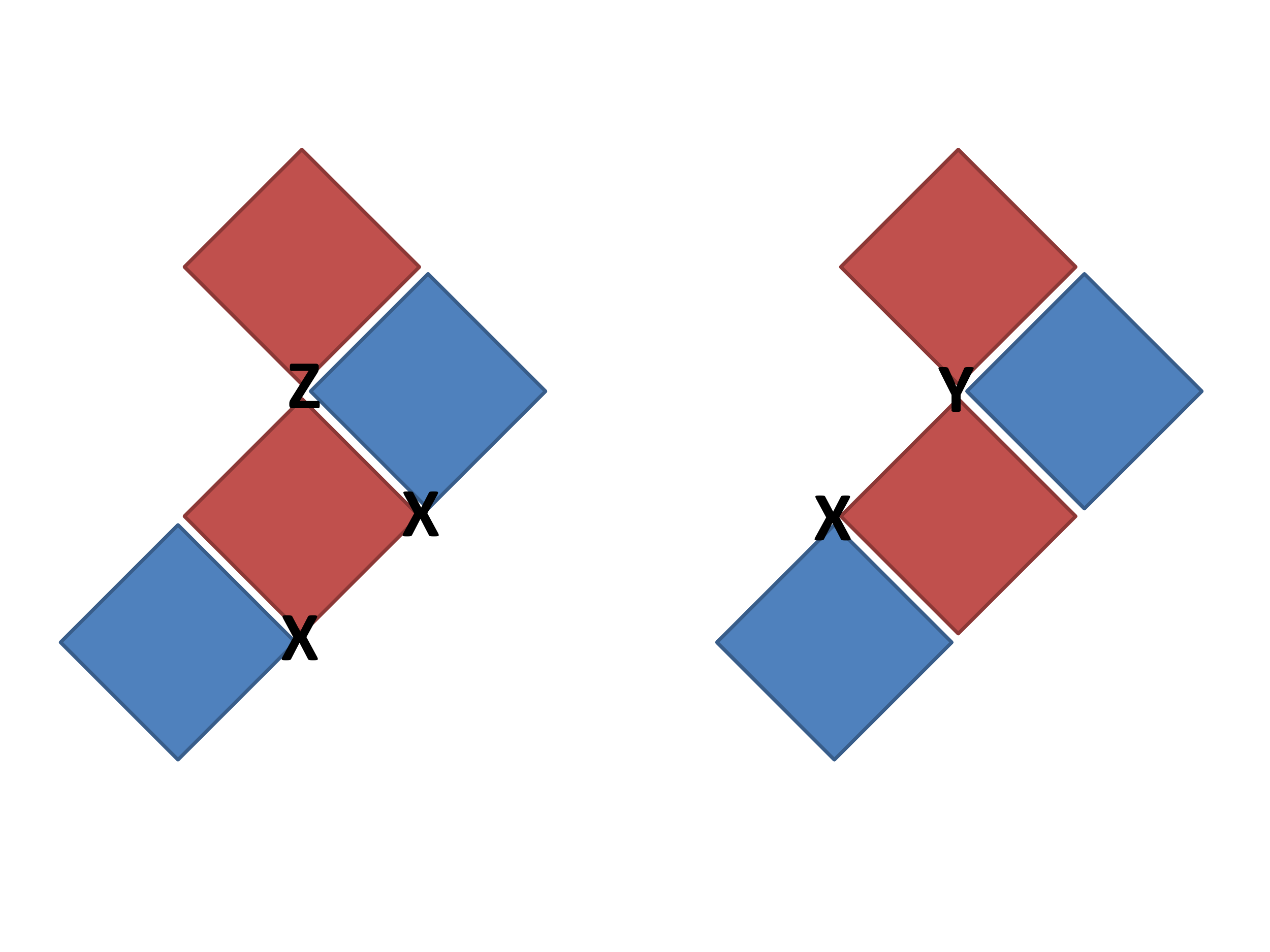}}
	\put(-0.09,0.28){\Large a)}
	\put(0.43,0.28){\Large b)}
  \end{picture}
  \caption{Squares represent stabilizers at which a $-1$ eigenvalue has been measured, i.e., anyons. Red stabilizers are tensor products of $\px$ and blue stabilizers are tensor products of $\pz$. 
  If the correlations between bit- and phase-flips present in depolarizing noise are ignored, both a) and b) are error patterns of minimal weight compatible with the anyon configuration.
  MWPM will thus result in either one of them with the same likelyhood.
  While enhanced MWPM will correctly assign a weight of $3$ to a) and a weight of $2$ to b), it considers only that of the two configurations which comes out of the matching algorithm.
  By contrast, if we start the single-temperature algorithm with configuration a), it will eventually apply the lower of the two red stabilizer operators and thereby convert it to the true minimum-weight configuration b).}
  \label{fig:minimalExample}
\end{figure}

Recall that we are not interested in the precise value of the integrals $\int_0^{\bar{\beta}}\m{d}\beta\left\langle n \right\rangle_{\beta, E}$, but only in knowing for which equivalence class $E$ this integral is smallest.
For this reason, calculating the whole integral is quite often an overkill. In fact, most of the relevant information contained in the function $\left\langle n \right\rangle_{\beta, E}$ can be extracted by finding its value for a \emph{single} inverse temperature $\beta^*$.

Assume that we determine the values $\left\langle n \right\rangle_{\beta^{*}, E}$ for some $\beta^{*}>0$ for all equivalence classes $E$. 
If the functions $\left\langle n \right\rangle_{\beta, E}$ for the different equivalence classes do not cross, knowing the values $\left\langle n \right\rangle_{\beta^{*}, E}$ is as good as knowing the whole integrals $\int_0^{\bar{\beta}}\m{d}\beta\left\langle n \right\rangle_{\beta, E}$
for deciding for which equivalence class $E$ the integral is smallest.

As $\beta\rightarrow0$, each qubit is affected by an $x$-, $y$-, or $z$-error or no error at all with probability $\frac{1}{4}$, 
so $\left\langle n \right\rangle_{\beta, E}\rightarrow\frac{3}{4}n_{\text{qubits}}$, where $n_{\text{qubits}}=n_{\text{stab}}+1$ is the number of data qubits in the code.
The low-$\beta$ tail of the function $\left\langle n \right\rangle_{\beta, E}$ thus contains almost no information about the equivalence class $E$.
So while the integral $\int_0^{\bar{\beta}}\m{d}\beta\left\langle n \right\rangle_{\beta, E}$ is dominated by its low-$\beta$ part ($\left\langle n \right\rangle_{\beta, E}$ is a monotonically decreasing function of $\beta$),
the \emph{differences} between these integrals for the different equivalence classes are mainly due to their high-$\beta$ part. 
So even if there are crossings in the low-$\beta$ tails of the functions $\left\langle n \right\rangle_{\beta, E}$, basing the decision for the most likely equivalence class on a single value $\left\langle n \right\rangle_{\beta^{*}, E}$ is likely to yield the same
outcome as basing the decision on the whole integral $\int_0^{\bar{\beta}}\m{d}\beta\left\langle n \right\rangle_{\beta, E}$.
We thus define our \emph{single-temperature algorithm} as sampling the values $\left\langle n \right\rangle_{\beta^{*}, E}$ for all equivalence classes and performing error correction in accordance with the equivalence class $E$ for which this value is smallest.

This algorithm has only two free parameters, namely $\beta^{*}$ and $n_{\m{sample}}$, the number of steps for which we perform the Metropolis algorithm in order to sample $\left\langle n \right\rangle_{\beta^{*}, E}$.
For $\beta^*\rightarrow\infty$ (zero temperature), the single-temperature algorithm will never increase the weight of an error configuration.
Still, it provides an improvement over enhanced MWPM since applying stabilizers allows to find error configurations which, taking correlations between bit- and phase-flips into account, are of lower weight than the ones found by MWPM.
This does not require that the weight of the error configuration be ever increased, see Fig.~\ref{fig:minimalExample} for an illustration.
At finite temperature, a second improvement over (enhanced) MWPM comes into play. 
Namely, temperature allows to take entropic contributions to the free energy into account, i.e., consider error configurations which are not of minimal weight but give a non-negligible contribution to the free energy due to their large number.
However, for $\beta^*\rightarrow0$ (infinite temperature) the single-temperature algorithm becomes useless ($\left\langle n \right\rangle_{\beta^{*}=0, E}=\frac{3}{4}n_{\text{qubits}}$ for all equivalence classes),
such that some finite value of $\beta^*$ is optimal.
Indeed, we find empirically that for depolarizing noise the optimal values for $\beta^{*}$ are close to $\bar{\beta}$.
We thus set $\beta^{*}=\bar{\beta}$ throughout for this error model.

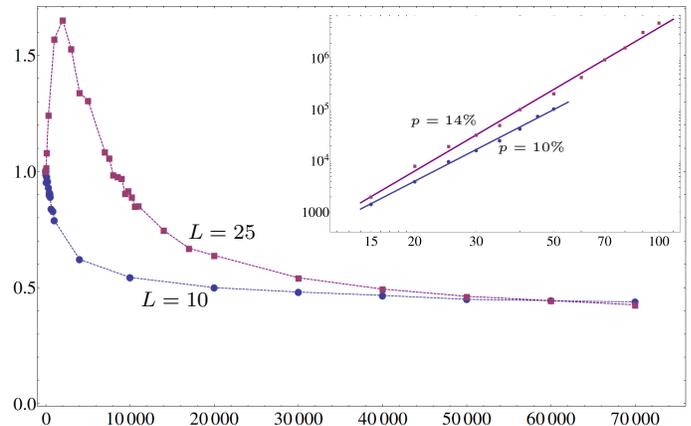
\begin{figure}
  \setlength{\unitlength}{0.5\textwidth}
  \begin{picture}(0.8,0.64)
	\put(-0.12,-0.02){\includegraphics[width=0.50\textwidth]{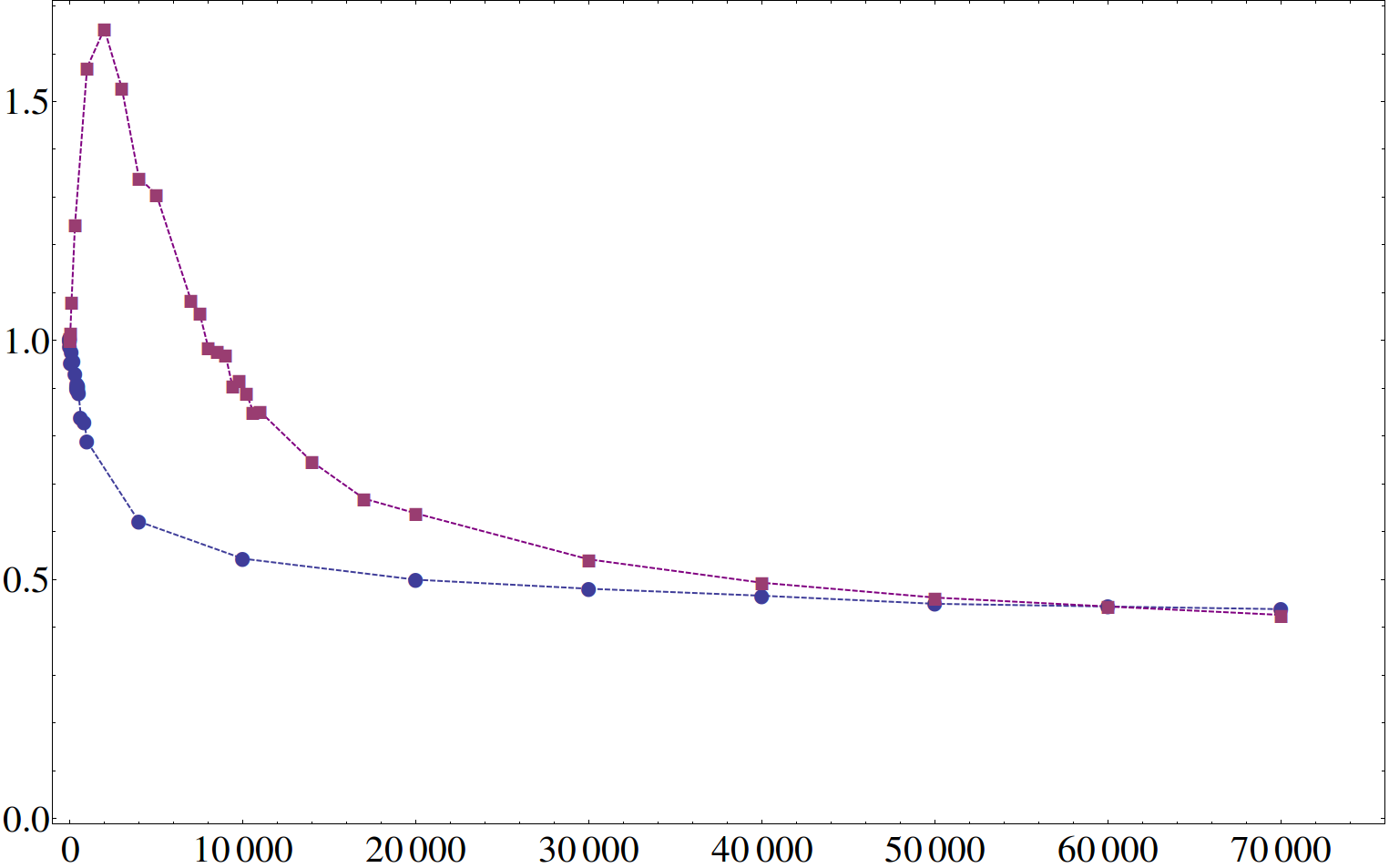}}
	\put(0.31,0.25){\includegraphics[width=0.28\textwidth]{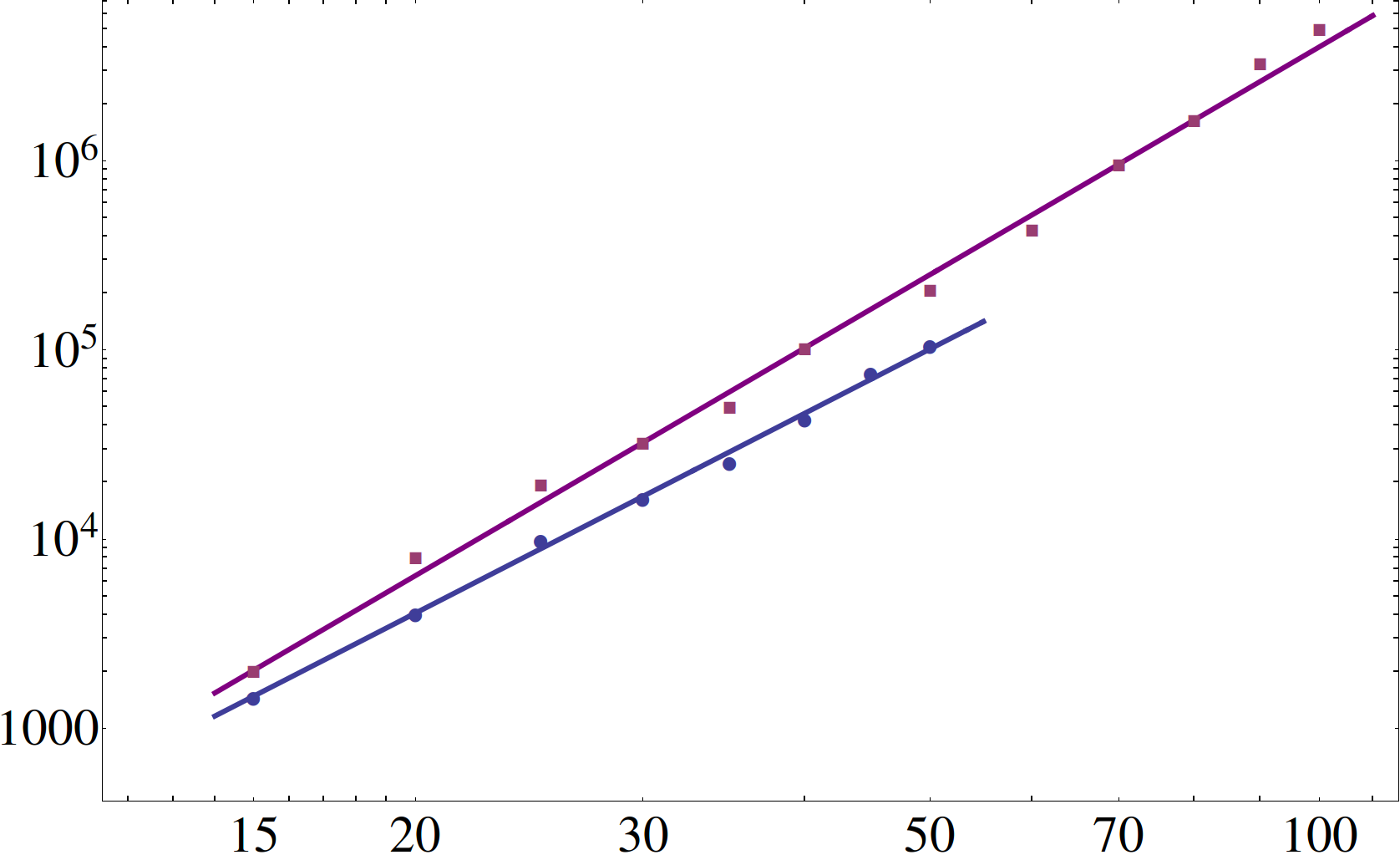}}

	\put(0.07,0.16){\footnotesize $L=10$}
	\put(0.14,0.26){\footnotesize $L=25$}

	\put(0.47,0.43){\tiny $p=14\%$}
	\put(0.60,0.39){\tiny $p=10\%$}

  \end{picture}
  \caption{
  The ratio of the logical error rate achieved with the single-temperature algorithm to the lower of the two error rates achieved by the two MWPM-algorithms as a function of $n_{\text{sample}}$.
  The data were obtained for a depolarization rate of $p=10\%$ and for code sizes $L=10$ and $L=25$, respectively.
  The inset shows the value of $n_{\text{sample}}$ which is necessary to achieve a unit ratio against $L$ for the case of depolarizing noise with $p=10\%$ and $p=14\%$. 
  The fitting lines correspond to the functions $0.11L^{3.51}$ and $0.04L^{4.00}$. 
  Each data point in the two figures is averaged over as many error configurations as were required for $2\,000$ logical errors to occur. 
  }
  \label{fig:depolAchieveMWPM}
\end{figure}

As for $n_{\m{sample}}$, we may ask how many Metropolis steps are necessary for our single-temperature algorithm to achieve logical error rates below those achievable with MWPM. In order to set the bar high, we compare our algorithm with the better of either standard or enhanced MWPM, i.e., that with the lower error rate.
Fig.~\ref{fig:depolAchieveMWPM} shows the ratio of the logical error rate achieved by the single-temperature algorithm divided by the smaller of the two logical error rates achievable by the two variants of MWPM.
If $L$ is higher than a certain $p$-dependent threshold, the logical error rate will be increased if $n_{\m{sample}}$ is too small, and only improve when it is made larger.
For some fixed $L$, this regime of increased logical error rate vanishes if $p$ is small enough, see the blue curve in Fig.~\ref{fig:depolAchieveMWPM}.
Then, already a handful of Metropolis steps is sufficient in order to outperform both variants of MWPM. 
As for the limit of a vanishing error rate $p$, recall that in this limit enhanced MWPM performs optimally and the single-temperature algorithm becomes redundant.

Let us discuss the scaling of the necessary values of $n_{\m{sample}}$ as a function of $L$ if we are in the regime where the single-temperature algorithm \emph{can} perform worse than (at least one variant of) MWPM when $n_{\m{sample}}$ is too low.
The approximation made by MWPM is, effectively, to approximate the free energy for each equivalence class by the number of errors in its minimum weight error chain. 
The simplest improvement to this that can be achieved by the single temperature algorithm is to calculate $\langle n \rangle_{\beta, E} $ by sampling within the vicinity of the minimum weight chains. 
This will give a better approximation of the free energy by taking into account some of the effects of entropy. 
Since such sampling requires only $O(1)$ deformations per string in the minimum weight error chain, this approximation is equivalent to assuming that the autocorrelation time for the calculation of $\langle n \rangle_{\beta, E} $ 
(when starting from the minimum weight chain) is $O(1)$ for each string. 
The runtime complexity required to generate independent Metropolis samples for the entire code is then $O(L^2)$,
which is thus the time-scale needed to estimate $\left\langle n \right\rangle_{\beta, E}$ up to some given \emph{relative} error.
The quantities $\left\langle n \right\rangle_{\beta, E}$ themselves grow like $O(L^2)$, and so does a constant relative error. 
However, the distinguishability, i.e., the difference in the quantity $\left\langle n \right\rangle_{\beta, E}$ between the correct and the remaining equivalence classes, only grows like $O(L)$ below threshold (see below), such that the relative difference between the equivalence classes \emph{decreases} like $O(L^{-1})$.
A constant relative error is thus not sufficient -- we need a relative error of order $O(L^{-1})$. As the relative error decreases with the inverse square root of the sample size, this leads to a further factor $O(L^2)$ in the runtime complexity.
The inset in Fig.~\ref{fig:depolAchieveMWPM} numerically verifies the $O(L^4)$ scaling anticipated from the above analysis.

\begin{figure}
  \setlength{\unitlength}{0.5\textwidth}
  \begin{picture}(0.8,0.7)
	\put(-0.05,0.00){\includegraphics[width=0.45\textwidth]{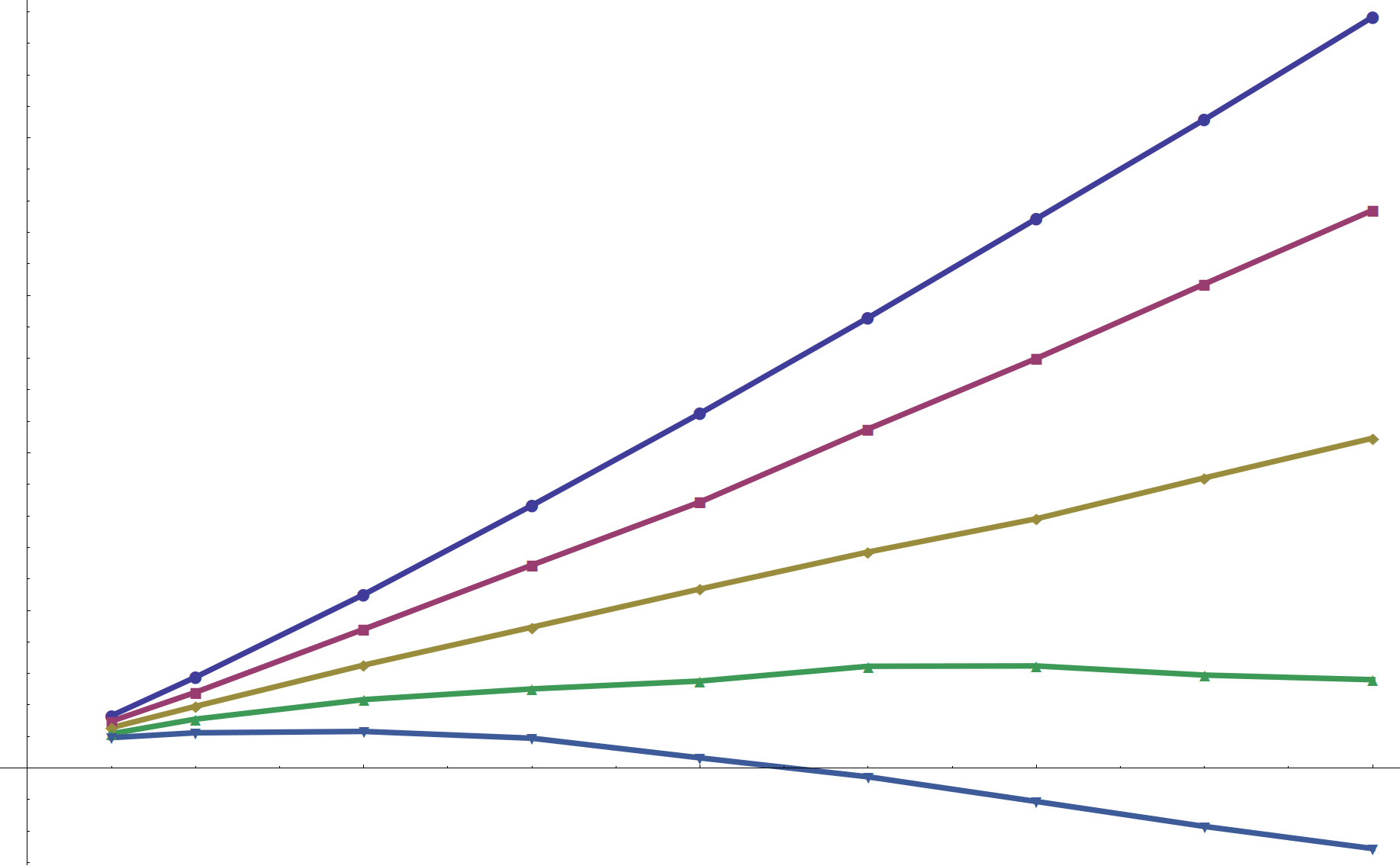}}
	\put(0.10,0.40){\includegraphics[width=0.16\textwidth]{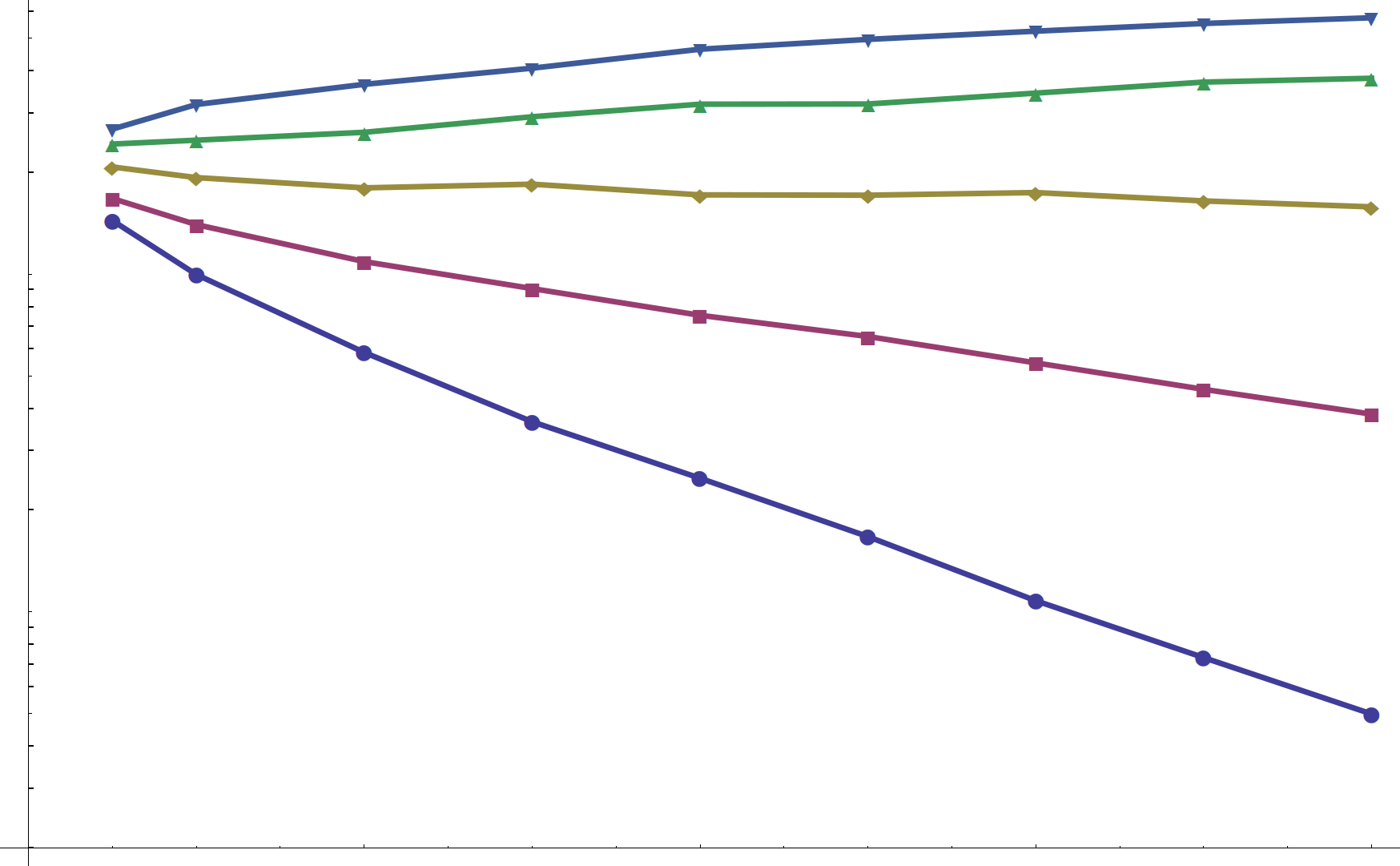}}
	\put(-0.13,0.13){\begin{sideways}$\min\left\langle n \right\rangle_{\beta^{*}, \text{false}}-\left\langle n \right\rangle_{\beta^{*}, \text{true}}$\end{sideways}}

	\put(0.06,0.03){\footnotesize $10$}
	\put(0.165,0.03){\footnotesize $20$}
	\put(0.275,0.03){\footnotesize $30$}
	\put(0.38,0.03){\footnotesize $40$}
	\put(0.49,0.03){\footnotesize $50$}
	\put(0.60,0.03){\footnotesize $60$}
	\put(0.705,0.03){\footnotesize $70$}
	\put(0.81,0.03){\footnotesize $80$}

	\put(-0.055,0.155){\footnotesize $5$}
	\put(-0.07,0.255){\footnotesize $10$}
	\put(-0.07,0.36){\footnotesize $15$}
	\put(-0.07,0.46){\footnotesize $20$}

	\put(0.131,0.385){\tiny $10$}
	\put(0.170,0.385){\tiny $20$}
	\put(0.209,0.385){\tiny $30$}
	\put(0.248,0.385){\tiny $40$}
	\put(0.286,0.385){\tiny $50$}
	\put(0.325,0.385){\tiny $60$}
	\put(0.364,0.385){\tiny $70$}
	\put(0.402,0.385){\tiny $80$}

	\put(0.042,0.422){\tiny $0.005$}
	\put(0.042,0.450){\tiny $0.010$}
	\put(0.042,0.506){\tiny $0.050$}
	\put(0.042,0.530){\tiny $0.100$}
	\put(0.042,0.590){\tiny $0.500$}

  \end{picture}
  \caption{The plot shows the distinguishabilities $\min\left\langle n \right\rangle_{\beta^{*}, \text{false}}-\left\langle n \right\rangle_{\beta^{*}, \text{true}}$ (vertical axis; we use $\beta^{*}=\bar{\beta}$ and $n_{\m{sample}}=L^4$) 
for various depolarization rates $p$ and code sizes $L$ (horizontal axis). 
Different lines correspond to different depolarization rates $p$. From top to bottom we have $p = 13\%, 14\%, 15\%, 16\%, 17\%$.
Each data point is averaged over as many error configurations as are necessary to obtain $2\,000$ logical errors.
The inset shows the logical error rates of our single-temperature algorithm for the same system sizes and depolarization rates, with depolarization rates increasing from bottom to top.
}
  \label{fig:depolFDiffs}
\end{figure}
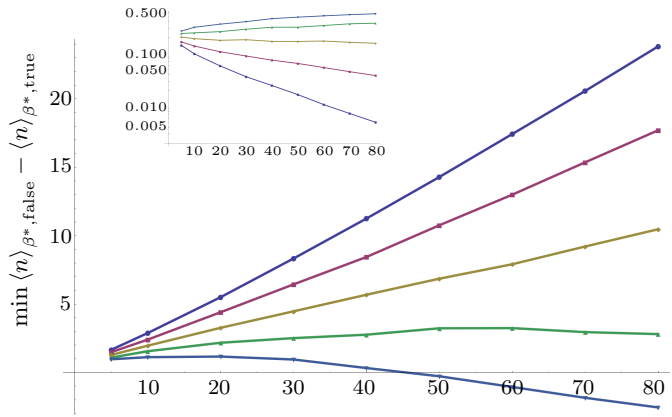

The quantity which determines whether error correction will be successful is the difference $\min\left\langle n \right\rangle_{\beta^{*}, \text{false}}-\left\langle n \right\rangle_{\beta^{*}, \text{true}}$,
where $\min\left\langle n \right\rangle_{\beta^{*}, \text{false}}$ denotes the minimal averaged number of errors of all three false equivalence classes. This difference is displayed for various values of $p$ and $L$ in Fig.~\ref{fig:depolFDiffs}.
For $p<16\%$ this difference increases linearly with $L$, while for $p=17\%$ it becomes even negative for large enough $L$. 
This is to be expected: for an error rate sufficiently close to the $18.9\%$ threshold, each of the averages $\left\langle n \right\rangle_{\beta^{*}, E}$ for the four equivalence classes $E$ has the same probability for being the smallest one, such that the probability that one of the three false equivalence classes becomes minimal approaches $\frac{3}{4}$.

The inset in Fig.~\ref{fig:depolFDiffs} shows the logical error rates achievable with our single-temperature algorithm if we set $n_{\m{sample}}=L^4$.
There is a threshold for $p$ between $15\%$ and $16\%$ below which the logical error rate decreases exponentially with $L$. 
This means that the threshold error rate for our algorithm is significantly below the theoretical maximum of $18.9\%$ \cite{Bombin12} and the value of $18.5\%$ achieved in Ref.~\cite{Woot12a} but closer to the threshold of MWPM \cite{Wang10}.
This is unsurprising since we use MWPM as a starting point for our Markov chains and use only the runtime complexity for $n_{\m{sample}}$ which is necessary to match MWPM.
However, the relevant figures of merit in practice are the logical error rates achievable well below threshold (where our algorithm offers significant improvement over MWPM, see below) and the runtime complexity (where our algorithm offers significant improvement over the algorithm of Ref.~\cite{Woot12a}).

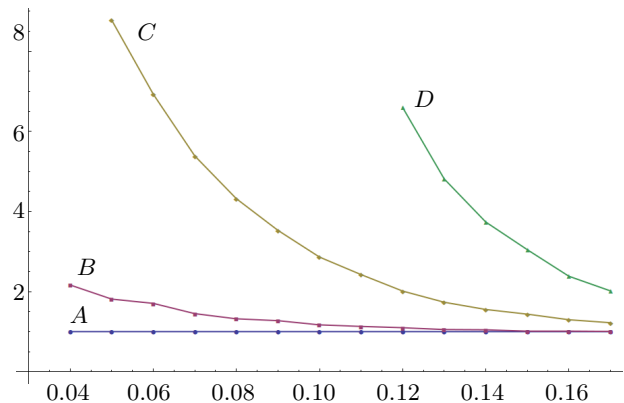
\begin{figure}
  \setlength{\unitlength}{0.5\textwidth}
  \begin{picture}(0.8,0.6)
	\put(-0.05,0.00){\includegraphics[width=0.45\textwidth]{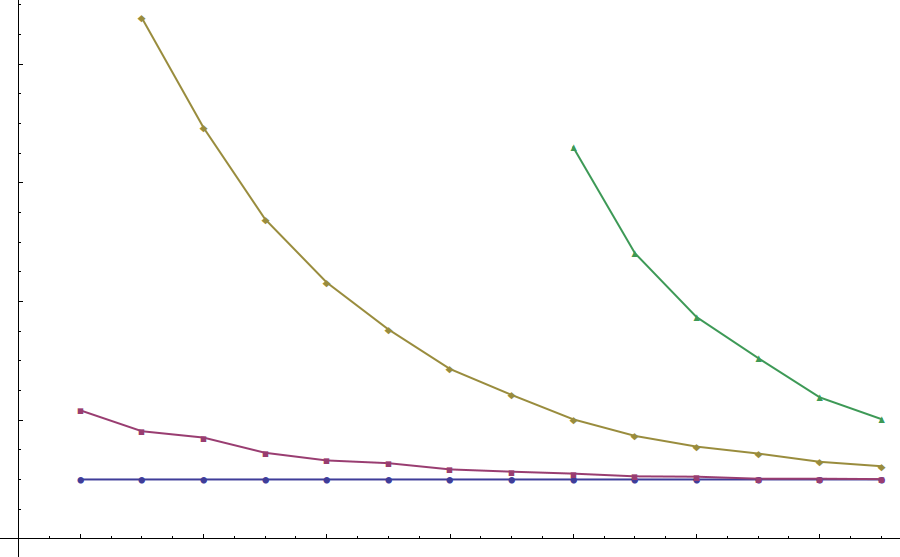}}
	
	\put(-0.005,-0.025){$0.04$}
	\put(0.115,-0.025){$0.06$}
	\put(0.24,-0.025){$0.08$}
	\put(0.36,-0.025){$0.10$}
	\put(0.485,-0.025){$0.12$}
	\put(0.61,-0.025){$0.14$}
	\put(0.735,-0.025){$0.16$}

	\put(-0.055,0.124){$2$}
	\put(-0.055,0.245){$4$}
	\put(-0.055,0.36){$6$}
	\put(-0.055,0.51){$8$}

	\put(0.03,0.09){$A$}
	\put(0.04,0.16){$B$}
	\put(0.13,0.51){$C$}
	\put(0.54,0.41){$D$}

  \end{picture}
  \caption{The plots show the logical error rate of MWPM (algorithm A in the main text) divided by the logical error rate of algorithms A to D described in the main text, for various depolarization rates $p$ (horizontal axis) and a code of linear size $L=15$.
Each data point is averaged over as many error configurations as are necessary to obtain $2\,000$ logical errors.
A value greater than $1$ denotes an increase in effectiveness over MWPM, with greater increases for higher values.}
  \label{fig:differentAlgos}
\end{figure}

Fig.~\ref{fig:differentAlgos} compares the logical error rates achievable with different algorithms for $L=15$ and different depolarization rates $p$. 
We set the logical error rates achievable with standard MWPM to unity and divide them by the logical error rates achievable with alternate algorithms. The algorithms which are displayed are:
\begin{itemize}
 \item[A)] Standard, unimproved MPWM, as employed in Ref.~\cite{Wang10}. An $x$- and a $z$-error on the same qubit count as two errors.
 \item[B)] Enhanced MWPM. An $x$- and a $z$-error on the same qubit count as two errors during the matching, but as one error during a final comparison of all equivalence classes. 
	    Corresponds to the single-temperature algorithm with $n_{\m{sample}}=0$.
 \item[C)] Single-temperature algorithm with $n_{\m{sample}}=L^4$ Metropolis steps and $\beta^{*}=\bar{\beta}$.
 \item[D)] The parallel-tempering algorithm developed in Ref.~\cite{Woot12a}.
\end{itemize}
The logical error rates decrease from A to D, while the runtime complexities increase.
We see that the advantages achievable over algorithm A vanish as $p\rightarrow p_c=18.9\%$ but increase the lower $p$ gets.
We believe our single-temperature algorithm C to offer the most attractive trade-off ratio between low logical error rates and low classical runtime complexity, 
as it increases the latter only modestly compared with algorithms A and B, while algorithm D has a super-polynomial (in $L$) runtime complexity.

We have verified numerically that estimating the entire integral $\int_0^{\bar{\beta}}\m{d}\beta\left\langle n \right\rangle_{\beta, E}$ by sampling the values $\left\langle n \right\rangle_{\beta, E}$ 
for $21$ equidistant temperatures $\beta$ over $L^4$ Metropolis steps and then applying Simpson's quadrature formula leads only to modest improvements over algorithm C.
It would be more beneficial to invest the additional computational cost into increasing $n_{\m{sample}}$ in algorithm C.

\begin{figure}
  \setlength{\unitlength}{0.5\textwidth}
  \begin{picture}(0.8,0.6)
	\put(-0.05,0.00){\includegraphics[width=0.45\textwidth]{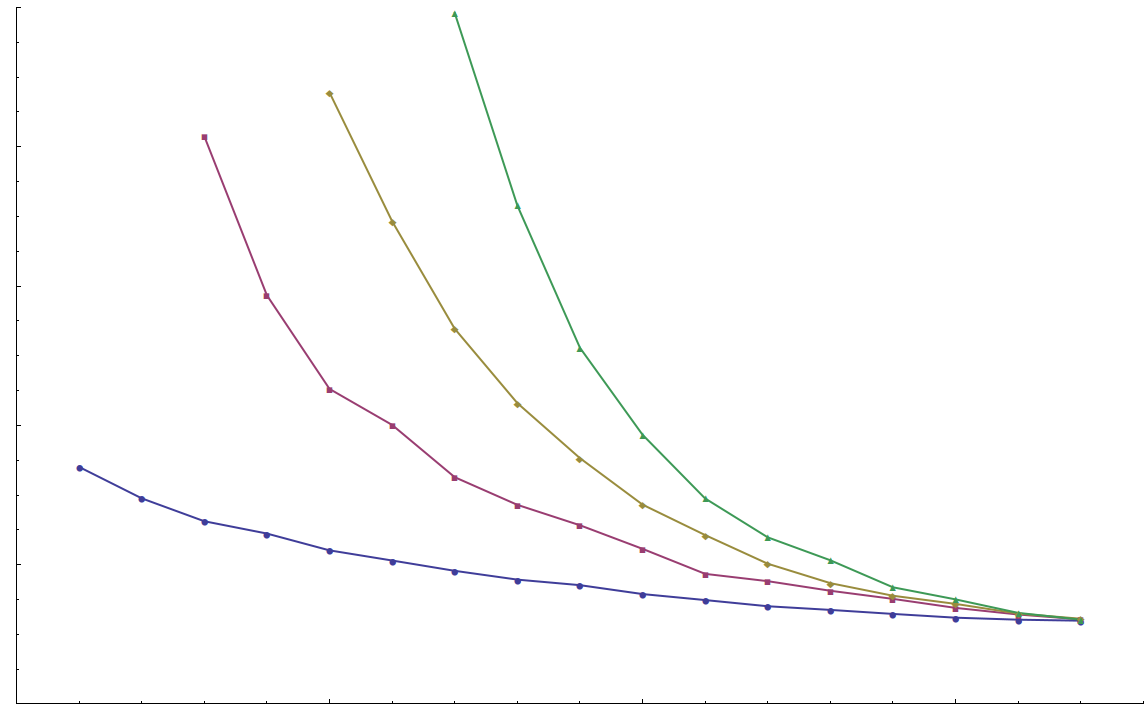}}

	\put(0.17,-0.025){$0.05$}
	\put(0.415,-0.025){$0.10$}
	\put(0.665,-0.025){$0.15$}

	\put(-0.065,0.11){$2$}
	\put(-0.065,0.22){$4$}
	\put(-0.065,0.33){$6$}
	\put(-0.065,0.44){$8$}
	\put(-0.081,0.55){$10$}

	\put(-0.01,0.21){$L=5$}
	\put(0.05,0.47){$L=10$}
	\put(0.17,0.50){$L=15$}
	\put(0.32,0.545){$L=25$}

  \end{picture}
  \caption{The plots show the logical error rate of algorithm A divided by the logical error rate of algorithm C for various depolarization rates $p$ (horizontal axis) and code sizes $L$.
Each data point is averaged over as many error configurations as are necessary to obtain $2\,000$ logical errors.}
  \label{fig:ST}
\end{figure}
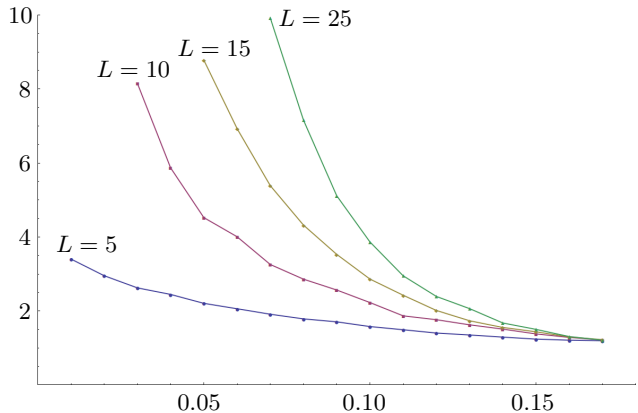

Fig.~\ref{fig:ST} shows the logical error rates of algorithm A divided by those of algorithm C for various values of $p$ and $L$. We see the advantage of algorithm C increasing for lower $p$ and larger $L$.
Note that we expect real quantum computers to be operated at error rates $p$ below and code distances $L$ above those displayed in Fig.~\ref{fig:ST}.

To get an idea about what effect this reduction of the logical error rates has on the necessary code size, let us have a look at the code size which is needed for a proof of principle experiment.
I.e., given some physical error rate $p$, which code size $L$ is needed to bring the logical error rate below the physical one?
For $p=13\%$, we need $L\geq6$ in order to achieve a logical error rate below $p$ with algorithm C and need $L\geq15$ with algorithm A.
So by modestly enhancing the classical runtime complexity, we are able to reduce the number of physical data qubits required for such a proof of principle experiment from $421$ to $61$.

The behavior of the curves in Figs.~\ref{fig:differentAlgos} and \ref{fig:ST} in the limit $p\rightarrow0$ demonstrates that the algorithms presented here do indeed achieve the predicted increased effectiveness over standard MWPM in the low $p$ limit. The ratio of the logical error rate of algorithm $A$ to the logical error rate of algorithms $B$ to $D$ can therefore be expected to be exponentially large in $L$ for $p\rightarrow0$. In this limit, the MCMC procedure becomes unnecessary and we can rely on algorithm $B$.

The advantage of algorithm C over algorithms A and B is due to at least two different reasons. First, MWPM is naturally suited to the error model of independent bit- and phase-flips, where each type of anyon can be treated independent of the other.
It is much less suited to error models such as depolarizing noise which feature correlations between bit- and phase-flip errors. Algorithm B can only partially overcome these limitations. 
Second, while algorithms A and B are based on finding the most likely error chain and hoping that it is an element of the most likely equivalence class, algorithms C to F are based on finding the most likely equivalence class.
An interesting question is thus how our single-temperature algorithm compares with MWPM for independent bit- and phase-flip errors, the error model MWPM is best suited to and where only the second advantage applies.
Empirically, we find that for this error model choosing $\beta^*=0.85\bar{\beta}$ works best and that for a bit-/phase-flip probability of $10\%$ (close to the theoretical threshold) $O(L^4)$ Metropolis steps are again sufficient to achieve a logical error rate below the one of MWPM \cite{MWPM_comment}, with $0.38L^{3.77}$ giving the best fit to the required number.
For $L=30$, a bit-/phase-flip probability of $8\%$, and by sampling over $10L^4$ Metropolis steps we achieve a logical error rate which is a factor $1.3$ lower than the one of MWPM.
So significant improvements over MWPM can be achieved even for the error model best suited to it, though the advantage is much more drastic for an error model with correlations between bit- and phase-flip errors, with which our single-temperature algorithm can deal very naturally.

\section{Parallelization}\label{sec:parallelization}

\begin{figure}
  \setlength{\unitlength}{0.5\textwidth}
  \begin{picture}(0.8,0.8)
	\put(-0.15,0.00){\includegraphics[width=0.55\textwidth]{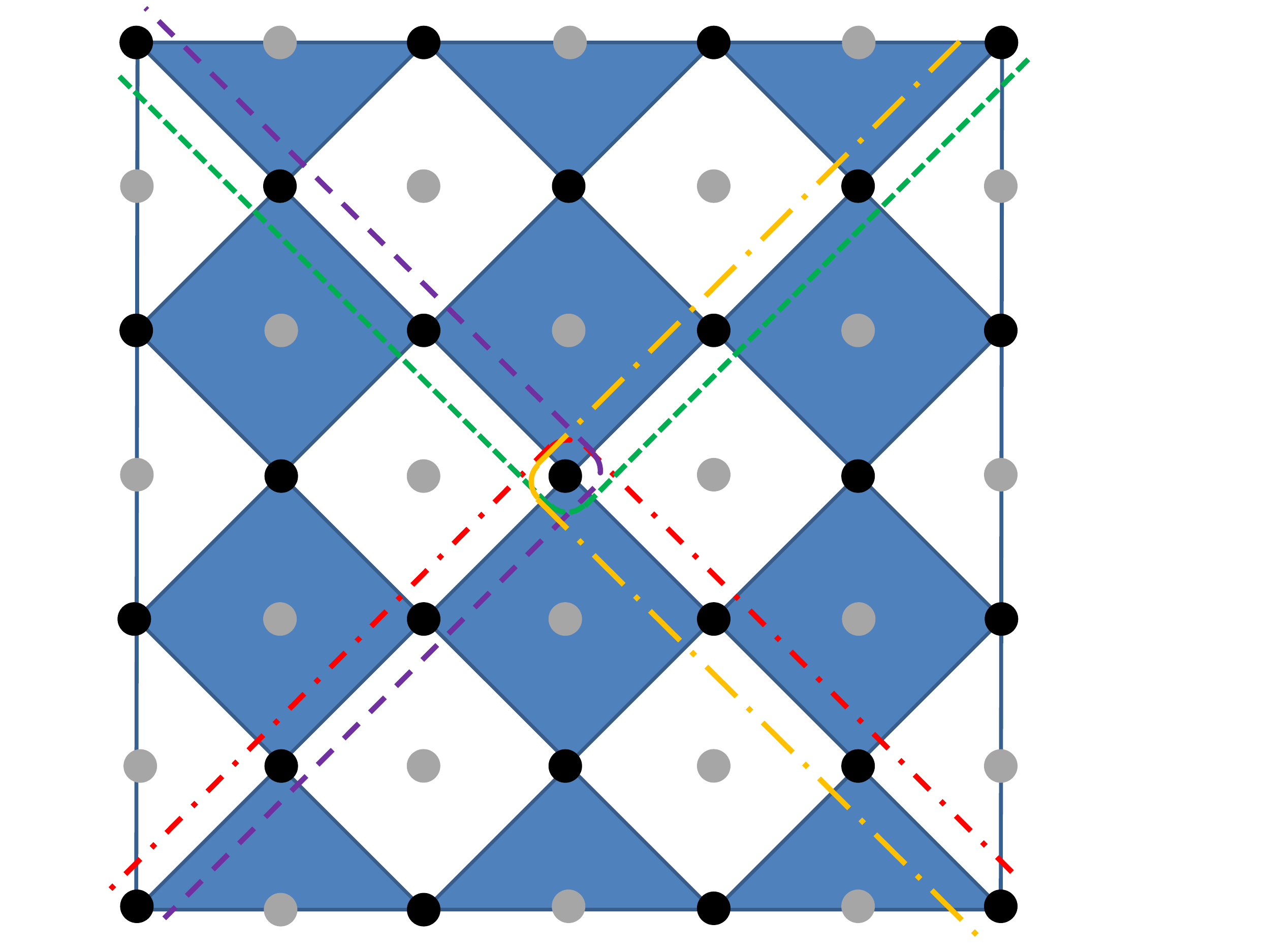}}
  \end{picture}
  \caption{The dashed lines partition the code into rectangles of size $O(1)$. Data qubits along the boundaries belong to two rectangles and data qubits in the corners to four.}
  \label{fig:parallel}
\end{figure}

The runtime of our algorithm can be reduced by a factor $O(L^2)$ using parallelization which exploits the fact that our algorithm needs local changes only.
We partition the whole code into $O(L^2)$ rectangles of area $O(L^0)$. Adjacent rectangles overlap along lines of data qubits, while qubits in the corners belong to four rectangles, see Fig.~\ref{fig:parallel}.
The rectangles are collected into four groups $0, 1, 2, 3$ such that the rectangles within one group have no overlapping qubits.
At step $i$ of the Metropolis Markov chain, we choose one stabilizer in each rectangle belonging to group $(i\!\mod4)$ at random and probe whether to apply it or not according to the Metropolis procedure.
This way we can guarantee that no data qubit is affected by more than one applied stabilizer at each step.
If a randomly chosen stabilizer is applied and flips a qubit which is shared with an other rectangle (other rectangles), this flip is communicated to the adjacent rectangle(-s).
For each rectangle, we add up the number of local errors $n$ over the different Metropolis steps and calculate the total average $\left\langle n \right\rangle_{\beta^{*}, E}$ in the end.
To compensate for double-(quadruple-)counting, errors on qubits along the boundary thereby have to be discounted by a factor $\frac{1}{2}$ and errors on qubits in the corners by a factor $\frac{1}{4}$.
As we probe now $O(L^2)$ stabilizers in each time step, the runtime reduces from $O(L^4)$ to $O(L^2)$.

\section{Imperfect stabilizer measurements}

\begin{figure}
  \setlength{\unitlength}{0.5\textwidth}
  \begin{picture}(0.8,0.8)
	\put(-0.15,0.00){\includegraphics[width=0.55\textwidth]{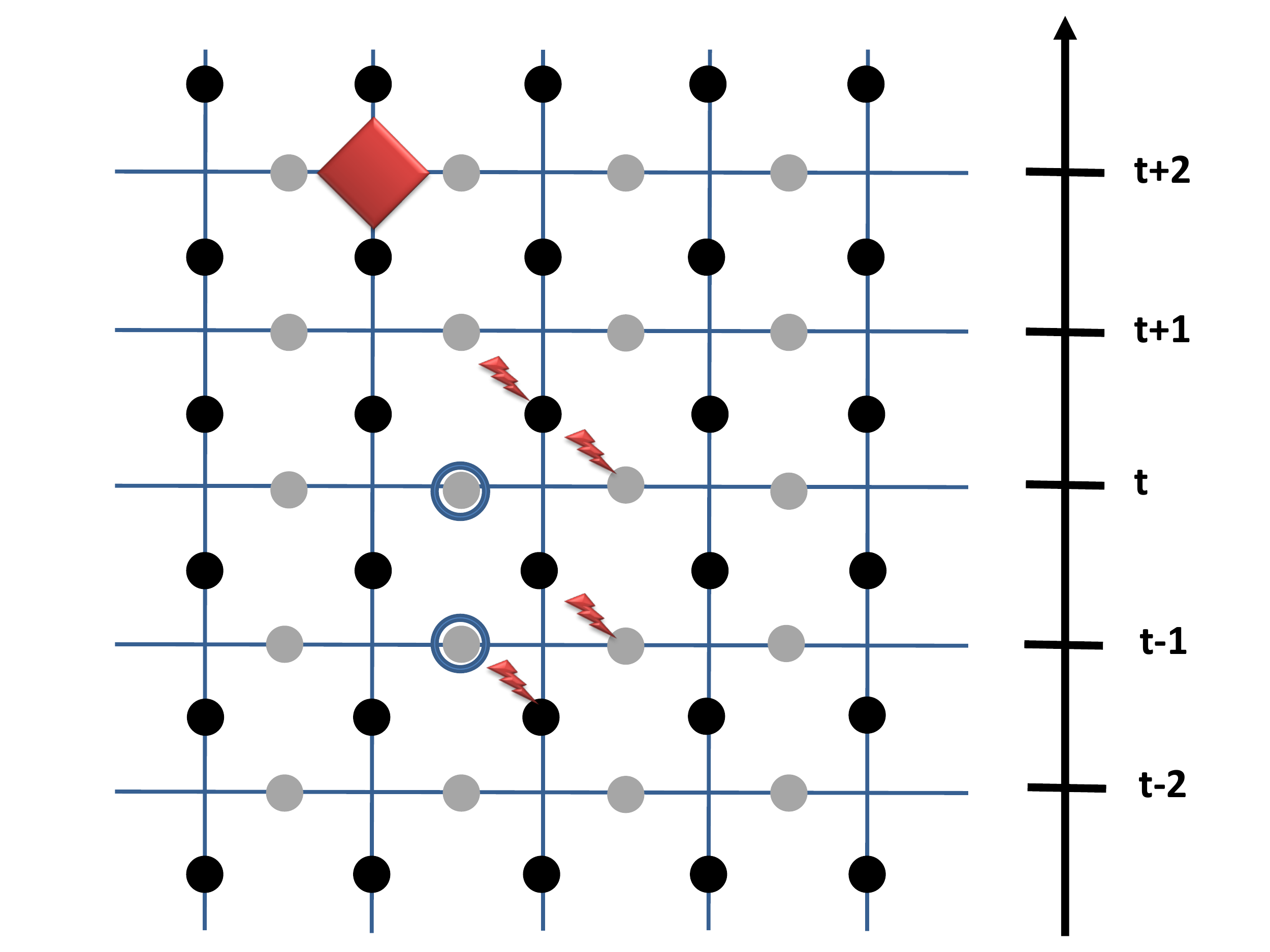}}
  \end{picture}
  \caption{Only one type of errors and stabilizers and a single chain of data qubits (black dots) and syndrome qubits (grey dots) are depicted here for simplicity. Time runs from bottom to top.
  Stabilizer operators are measured at times  $\ldots,t-1,t,t+1,\ldots$. The two encircled syndrome qubits have detected errors ($-1$ eigenvalues). A possible hypothesis is that no data qubit has suffered an error and both syndrome measurements have been erroneous.
  An alternate hypotheses would state that the data qubits indicated by red lightning bolts have suffered errors and that syndrome measurement by use of the syndrome qubits indicated by red lightning bolts has been erroneous. 
  Different equivalent hypotheses can be deformed into each other through the application of operators like the red square that invert whether a hypothetical error has happened or not at two adjacent syndrome qubits and one data qubit at two subsequent times.
}
  \label{fig:noisyStabilizer}
\end{figure}

Let us assume that each stabilizer measurement yields the wrong result with probability $p_M$. 
If stabilizers are measured by use of CNOT gates, this model is a simplification in that it ignores correlations between spatio-temporally nearby syndrome measurement failures induced by these gates \cite{Wang11}.
In order to make the failure probability small despite non-negligible $p_M$, we now necessarily need to perform stabilizer measurements at several times $t=1 , 2 , \ldots , t_{\max}$.
A hypothesis about what errors have happened then not only has to state which data qubit has suffered an error in which time interval $[t,t+1]$, but also which stabilizer measurement has been erroneous at which time $t$.
Such hypotheses can be deformed into equivalent ones by applying a bit-/phase-flip $\px$/$\pz$ to a particular qubit at time intervals $[t-1,t]$ and $[t,t+1]$ and inverting the hypothesis about whether the stabilizer measurements at time $t$ that anti-commute with this error have been erroneous 
(see the illustration in Fig.~\ref{fig:noisyStabilizer}).

In the case of depolarizing noise, a hypothesis that involves $n$ data-qubit errors and $m$ erroneous syndrome measurements has a relative probability
\begin{align}
 \left(\frac{p/3}{1-p}\right)^n\times\left(\frac{p_M}{1-p_M}\right)^m \equiv \exp\left[-\bar{\beta}(n+\xi m)\right]\ ,
\end{align}
where $\bar{\beta}$ is as defined in Eq.~(\ref{eq:betaBar}) and 
\begin{align}
 \xi = \frac{1}{\bar{\beta}}\log\frac{1-p_M}{p_M}\ .
\end{align}
The ``energy'' of a hypothesis is thus given by $n+\xi m$ where $\xi$ determines the relative weight of erroneous syndrome measurements to data qubit errors.
Our method to find the most probable equivalence class in the case of perfect stabilizer measurements can thus be generalized to the case where syndrome measurements fail with a considerable probability. 
Numerical results for the latter case will appear in future work.

\section{Conclusions}
We have developped two novel error correction algorithms -- enhanced MWPM and the single-temperature MCMC algorithm -- and compared them with each other and with standard MWPM over several regimes of the error rate $p$: 
close to threshold, intermediate values, and vanishing values. For the first two regimes, numerical simulations have provided us with insight into their respective performance, while in the third regime, analytical arguments are both unavoidable and possible.

The relevant regime in practice will be the second one. 
In this regime, no increase in the runtime complexity is necessary for the single-temperature algorithm to achieve lower logical error rates than the other two algorithms, 
so any CPU-time not needed to perform MWPM can be used to lower the logical error rate by the method described in this work.
We have numerically investigated the decreases in the logical error rates which are achievable if we are willing to increase the classical runtime complexity by $O(L^2)$.
Nothe that advantages much higher than those displayed Figs.~\ref{fig:differentAlgos} and Fig.~\ref{fig:ST} can likely be achieved, as we have probed only relatively small values of $L$ and high values of $p$, 
quantum computers will be operated at error rates significantly below threshold, and the advantages increase for higher values of $L$ and lower values of $p$.

Besides lowering the logical error rate for a given code size, our algorithm also reduces the code size necessary for a proof of principle experiment and thus reduces the experimental requirements for such an experiment.

Like the renormalization group method \cite{Ducl13} but unlike MWPM our algorithm readily generalizes to the $\mathbb{Z}_d$ toric codes with $d>2$ \cite{Kitaev03}.
Our algorithm relies on the availability of a low energy state of each equivalence class as a starting point for the Markov chains.
In the case of $d=2$ (studied in this work), such a low energy state can be efficiently obtained by use of MWPM, while for $d>2$ the Broom algorithm of Ref.~\cite{Brav11} may be applied.
How starting the Markov chains with a random (and thus likely high energy) input state from each equivalence state would affect the necessary runtime remains unclear and will be the subject of future work.

\section{Acknowledgments}

This work is supported by the Swiss NSF, NCCR QSIT, and IARPA.
We thank Austin Fowler for encouraging us to discuss the $p\rightarrow0$ limit and David Poulin for comments concerning the applicability of our algorithm.



\begin{thebibliography}{38}
\bibitem{comparison_comment}
As the error correction algorithm of Ref.~\cite{Ducl09} itself allows for trade-offs between runtime and error rates, and MWPM based algorithms have attained most interest in the literature, we do not directly compare our algorithm to the former.
\bibitem{log_comment}
Neglecting, again, $O(\log L)$ factors due to the random number generator which is necessary to propose one of the $O(L^2)$ stabilizer operators for the Metropolis procedure.
\bibitem{MWPM_comment}
Note that for independent bit- and phase-flip errors, there is only one variant of the MWPM algorithm, as different error types are by assumption uncorrelated.
\end{thebibliography}
\end{document}